**Affects of Electrical Image Potentials and Solvation Energy on Salt Ions Near a Metallic or Dielectric Wall**

J. B. Sokoloff[1,2], 1.Physics Department, Florida Atlantic University, Boca Raton, Florida 33431; 2. Physics Department and Center for Interdisciplinary Research in Complex Systems, Northeastern University, Boston, MA 02115

Electrical image potentials near a metallic or a dielectric wall of higher dielectric constant than that of the solution are attractive, and therefore, could concentrate salt ions near the wall. In fact, ions in room temperature ionic liquids have been observed to precipitate near a metallic surface (but not near a nonmetallic surface). It will be argued that a likely reason for why precipitation of ions in salt water due to electrical image forces has not as yet been observed is that the solvation of the ions is reduced near the wall. This results in an energy barrier. This reduction occurs because of the large decrease near the wall of the dielectric constant of water normal to the wall. The conditions under which ions are able to get past the resulting energy barrier and concentrate at a solid wall, either as a result of a reduction of this barrier due to screening at high ion concentration or as a result of thermal activation over the barrier will be explored.

## I.    Introduction

There have in recent years been many studies of the behavior of salt ions near surfaces and interfaces[1-5]. Such studies have application in areas such as water desalination and purification[6-8], blue energy harvesting[9], supercapacitors[10,11], electrowetting[12] and ion absorption onto biological molecules[13] for example. Onsager and Samaras[14] have shown that electrical image charge forces acting on ions dissolved in water can explain the existence of an increase of the surface tension due to the presence of dissolved ions. Electrical image charge forces at the interface of a salt solution and air are repulsive. At the interface between the salt solution and a dielectric of higher dielectric constant than that of the solution or at the surface of a metal, in contrast, the force is attractive. It has also recently been shown that electrical image forces may play a role in the attraction of ions to the electrodes in capacitive desalination[1]. It has been suggested that electrical image forces near a metallic surface or a dielectric surface of high enough dielectric constant can result in a high concentration of salt ions at the surface, which could be used to desalinate salt water, as long as the surface does not have an oxide coating (on a metallic surface) that is thicker than a few Angstroms, as this would prevent the ions from getting close enough to the metallic material to feel the image potential[15]. This result is supported by a one-loop approximation variational calculation, which is the Debye-Huckel approximation[16]. In fact, precipitation of ions at a metallic surface from room temperature ionic liquids has been observed[2]. This raises the question of why this effect has not been observed in solutions of salt in water. One possibility is that most metals have an oxide coating. One conducting material that does not have an oxide coating is graphene, but pure graphene is a semimetal, and hence the electronic screening length is infinite[17]. As discussed in appendix A, however, the width of the image charge potential well is of the order of the screening length[18]. Consequently, the ions will not be concentrated near an undoped graphene wall. Graphene which is doped so that it behaves as a metal, however, could have an image potential well of finite width. Another possible explanation for why concentration of ions by electrical image forces has not been observed is the existence of a potential energy barrier near a solid surface, resulting from the reduction of ion solvation as a result of the large decrease of the dielectric constant of water normal to a solid wall within a distance $H = 0.75nm$ of the wall[19]. The existence of this energy barrier and possible ways that it can be overcome will be explored here. In section II, previous simple treatments of this problem are briefly summarized. In section III, the effects of the tensor nature of the permittivity of water near a solid wall are discussed. In section IV, it is shown that ionic screening can reduce the solvation energy sufficiently for highly concentrated salt solutions to make it possible for electrical image potentials to concentrate salt ions near a metallic wall. In section V thermal activation of ions over the solvation energy barrier is discussed.

## II. Summary of Simple Treatments of the Image Potential and Self-Energy

The self-energy of an ion is its electric field energy. Its solvation energy is equal to the difference between its self-energy when it is immersed in the solvent and when it is not immersed in the solvent. Let us consider the calculation of the solvation energy by calculating an ion's self-energy by direct integration, as was done in Ref. 3. Consider an interface of zero width located at z=0, separating a region with z>0, in which the dielectric constant is $\varepsilon_1$, from a region with z<0 which has a dielectric constant $\varepsilon_2$, with $\varepsilon_2 > \varepsilon_1$. Consider an ion of charge Q and radius $a$ whose center is located at $z = z_0 > a$. Then, following Ref. 3, its electrostatic energy is given by

$$U_{tot} = (1/2)\int d^3r \rho(\vec{r}) \int d^3r' \left[ \frac{1}{4\pi\varepsilon_1 |\vec{r}-\vec{r}'|} + \frac{\Delta_{12}}{4\pi\varepsilon_1 |\vec{r}-\vec{r}^*|} \right] \rho(\vec{r}'), \qquad (1)$$

where $\vec{r}^* = (x', y', -z')$, $\Delta_{12} = (\varepsilon_1 - \varepsilon_2)/(\varepsilon_1 + \varepsilon_2)$ and $\rho(\vec{r})$ is the charge density in the ion. If instead of a dielectric medium for $z < 0$ there is a metallic surface at $z = 0$, $\Delta_{12} = -1$. The integrals over the second term in the bracket is just the image charge potential energy for the ion. Since we are restricting the ion to $z_0 > a$, the second term is equal to the image charge potential energy for a point charge located at $\vec{r} = (0, 0, z_0)$. The first term in the square bracket in Eq. (1) is the Coulomb potential energy, which is independent of the presence of the wall or interface, and the second term (the image potential) accounts for the boundary conditions at the wall or interface. Therefore, the integral over the first term in the square bracket gives what would be the interaction of the ion's charge with itself in the absence of the wall or interface. This must be equal to the electric field energy in the absence of the wall or interface, which is equal to the integral of the electric field energy density $(1/2)\vec{D}\cdot\vec{E}$ over all volume, with $\vec{E}$ and $\vec{D}$ calculated from the Coulomb potential term alone, as if there were no wall or interface.

Let us now do the integrals over the first term in the brackets, assuming that the ion's net charge resides on its surface (a commonly used model for an ion's charge distribution). The potential at $\vec{r} = a\hat{z}$, where $\hat{z}$ is the unit vector in the z-direction, is found by doing the integral over $\vec{r}'$ on the surface of the sphere as illustrated below:

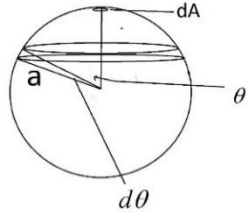

Figure 1: This figure illustrates the way that the integral in Eq. (2) was performed. The azimuthal angle is denoted by $\theta$ and $dA$ is an element of surface area located at $\vec{r} = a\hat{z}$.

The potential at $\vec{r} = a\hat{z}$ is given by

$$V = \int \frac{\sigma dA'}{4\pi\varepsilon_1 |\vec{r}-\vec{r}'|} = \frac{1}{4\pi\varepsilon_1} \int_0^\pi d\theta \frac{2\pi a^2 \sigma \sin\theta}{a[2(1-\cos\theta)]^{1/2}} = \frac{4\pi a^2 \sigma}{4\pi\varepsilon_1 a} = \frac{Q}{4\pi\varepsilon_1 a}, \qquad (2)$$

where Q is the charge of the ion, $\sigma = Q/(4\pi a^2)$, $dA' = 2\pi a^2 \sin\theta d\theta$ and $|\vec{r}-\vec{r}'| = \left[(a-a\cos\theta)^2 + (a\sin\theta)^2\right]^{1/2}$, where $\theta$ is the azimuthal angle. By symmetry, the potential is constant over the surface of the spherical shell. Therefore, the solvation energy is just the product of $Q$ and this potential times 1/2, or

$$U = \frac{Q^2}{8\pi\varepsilon_1 a}, \qquad (3)$$

the Born self-energy[20].

At least for the spherical shell, it is correct to use the dielectric constant outside the sphere when calculating the interaction between charge on different sides of the sphere, because of the following argument: Inside the sphere, the permittivity is just the permittivity of free space $\varepsilon_0$, but the net surface charge on the sphere is equal to the surface free charge Q minus the surface polarization charge, which is equal to $4\pi a^2 P$, where P is the polarization vector (i.e., the polarization per unit volume). The polarization P is given by

$$P = (\varepsilon_1 - \varepsilon_0)E = (\varepsilon_1 - \varepsilon_0)\frac{Q}{4\pi\varepsilon_1 a^2}. \qquad (4)$$

Then, the net surface charge is given by

$$Q - 4\pi a^2 P = \frac{\varepsilon_0}{\varepsilon_1} Q_{net}. \qquad (5)$$

Therefore, although pairs of elements of charge on different parts of the sphere interact with the Green's function

$$G(\vec{r}, \vec{r}\,') = \frac{1}{4\pi\varepsilon_0 |\vec{r} - \vec{r}\,'|}, \qquad (6)$$

since the actual net charge (the free charge minus the polarization charge) is equal to $(\varepsilon_0/\varepsilon_1)Q$, the effective G is Eq. (6) with $\varepsilon_1$ in place of $\varepsilon_0$ in its denominator.

### III. Treatment of the Self-Energy and Image Potential for the Tensor Permittivity Near a Wall

The permittivity of water within a distance $H$ from a surface is a tensor[4,21], which when diagonalized has diagonal elements $(\varepsilon_\|, \varepsilon_\|, \varepsilon_{perp})$. The element $\varepsilon_{perp} = 2.1\varepsilon_0$, in MKS units, where $\varepsilon_0$ is the permittivity of free space, and $\varepsilon_\|$ is much larger. The likely reason that the dielectric constant for a field directed normal to a solid wall is only equal to 2.1 [4,19] is that the presence of the wall inhibits the rotation of the dipoles of the water molecules located there away from and towards the wall. For fields that are parallel to the wall, however, there are no such restrictions. Therefore, the dielectric constant for fields parallel to the wall should be much larger, possibly as large of the bulk value of 81. The image charge potential and self-energy will now be calculated with this tensor permittivity. The potential for a point charge in a medium with the above tensor permittivity is the solution to the following Poisson equation in the region $z>0$,:

$$\varepsilon_{perp}\frac{\partial^2 \phi}{\partial z^2} + \varepsilon_\| \left( \frac{\partial^2 \phi}{\partial x^2} + \frac{\partial^2 \phi}{\partial y^2} \right) = -Q\delta(\vec{r} - \vec{r}_0), \qquad (7)$$

with the point charge $Q$ located at $\vec{r}_0 = z_0 \hat{z}$. If we make the substitution $x = \varepsilon_\|^{1/2} x'$, $y = \varepsilon_\|^{1/2} y'$, $z = \varepsilon_{perp}^{1/2} z'$ in Eq. (7), it becomes

$$\nabla'^2 \phi = -\frac{Q}{4\pi\varepsilon_\| \varepsilon_{perp}^{1/2}} \delta(x')\delta(y')\delta(z'-z_0'), \qquad (8)$$

where the operator $\nabla'^2 = \partial^2/\partial x'^2 + \partial^2/\partial y'^2 + \partial^2/\partial z'^2$ and where $z_0' = z_0/\varepsilon_{perp}^{1/2}$. Its solution is

$$\phi = \frac{Q}{4\pi\varepsilon_\| \varepsilon_{perp}^{1/2} [x'^2 + y'^2 + (z'-z_0')^2]^{1/2}} = \frac{Q}{4\pi\varepsilon_\|} \frac{1}{[(z-z_0)^2 + (\varepsilon_{perp}/\varepsilon_\|)r_\|^2]^{1/2}} \qquad (9)$$

where $\vec{r}_\| = x\hat{x} + y\hat{y}$. Then, the E-field is given by

$$\vec{E} = -\nabla\phi = \frac{Q}{4\pi\varepsilon_\|} \frac{(z-z_0)\hat{z} + (\varepsilon_{perp}/\varepsilon_\|)(x\hat{x} + y\hat{y})}{[(z-z_0)^2 + (\varepsilon_{perp}/\varepsilon_\|)(x^2+y^2)]^{3/2}}. \qquad (10)$$

Although this field was calculated for a point charge, as we will show below, it is also the electric field that would be produced by a charged spherical shell of radius $a$, whose surface charge distribution will now be determined. This model is introduced here because it will be used in section IV to illustrate the effect of ionic screening on the self-energy, since the self-energy can be calculated analytically for this model. The displacement field is given by

$$\vec{D} = -\vec{\varepsilon} \cdot \nabla \phi = \frac{Q}{4\pi}\left(\frac{\varepsilon_{perp}}{\varepsilon_{\|}}\right)\frac{(z-z_0)\hat{z} + x\hat{x} + y\hat{y}}{[(z-z_0)^2 + (\varepsilon_{perp}/\varepsilon_{\|})(x^2+y^2)]^{3/2}}, \qquad (11)$$

and hence, the surface charge density on the spherical shell of radius $a$ is given by

$$\sigma = \hat{r} \cdot \vec{D} = \frac{Q}{4\pi a^2}\left(\frac{\varepsilon_{perp}}{\varepsilon_{\|}}\right)\frac{1}{[\cos^2\theta + (\varepsilon_{perp}/\varepsilon_{\|})\sin^2\theta]^{3/2}}. \qquad (12)$$

For simplicity, we have set $z_0 = 0$ in Eq. (11), and $x, y$ and $z$ lying on the surface of the ion have been expressed in spherical coordinates, where $\theta$ is the azimuthal angle. It is easy to check by doing the integral in Eq. (12) that

$$Q = 2\pi a^2 \left(\frac{\varepsilon_{perp}}{\varepsilon_{\|}}\right)\frac{Q}{4\pi a^2}\int_0^\pi \frac{\sin\theta \, d\theta}{[\cos^2\theta + (\varepsilon_{perp}/\varepsilon_{\|})\sin^2\theta]^{3/2}}. \qquad (13)$$

The self-energy for the model described by Eqs. (9-13) is given by

$$U = (1/2)\int d^3r \vec{E} \cdot \vec{D} = (1/2)(2\pi)\frac{Q^2 \varepsilon_{perp}}{(4\pi)^2 \varepsilon_{\|}^2}\int_0^\pi d\theta \sin\theta \frac{1}{[\cos^2\theta + (\varepsilon_{perp}/\varepsilon_{\|})\sin^2\theta]^2}\int_a^\infty \frac{r^2 dr}{r^4}$$

$$= \frac{Q^2}{16\pi\varepsilon_{\|}a}\left[1 + (\varepsilon_{\|}/\varepsilon_{perp})(\varepsilon_{\|}/\varepsilon_{perp}-1)^{-1/2}\arctan(\varepsilon_{\|}/\varepsilon_{perp}-1)^{1/2}\right]. \qquad (14)$$

For $\varepsilon_{\|} = \varepsilon_{perp} = \varepsilon_1$, Eq. (14) reduces to Eq. (3), the Born expression for the self-energy. For $\varepsilon_{\|} > \varepsilon_{perp}$, it is larger than this value.

If we use the fact that $\varepsilon_{perp}/\varepsilon_{\|}$ is small compared to 1, we can calculate the self-energy for a uniformly charged spherical shell by direct integration partially analytically, as follows: Using the electrical potential for a tensor permittivity [i.e., Eq. (9)], the electrical potential at a point on the sphere denoted by $(\vec{r}_{\|}, z)$ is given by

$$\phi = \frac{\sigma}{4\pi\varepsilon_{\|}^{1/2}}\int ds' \oint d\ell'_{\|} \frac{1}{[(z'-z)^2 + (\varepsilon_{perp}/\varepsilon_{\|})|\vec{r}_{\|}'-\vec{r}_{\|}|^2]^{1/2}}, \qquad (15)$$

where $d\ell'_{\|}$ is the element of length around a circular path normal to the z-axis around the sphere, $ds'$ is the length of an arc on the surface of the sphere normal to the circular path and $\sigma = Q/(4\pi a^2)$. It is assumed here that it is correct to use the tensor permittivity valid outside the ion, which was shown in the last section to be correct for a scalar permittivity. The calculation is illustrated in Fig. 2.

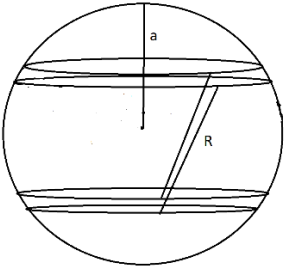

Fig. 2: This is an illustration of the integral in Eq. (15), where $R = [(z'-z)^2 + (\varepsilon_{perp}/\varepsilon_{\parallel})|\vec{r}_{\parallel}'-\vec{r}_{\parallel}|^2]^{1/2}$.

For simplicity, we evaluate the above integral for $\vec{r}_{\parallel} = 0$, because as we will argue below, this only introduces a negligibly small error. Using spherical coordinates, Eq. (15) becomes

$$\phi = \frac{\sigma}{4\pi\varepsilon_{\parallel}^{1/2}} \int_0^{\pi} \frac{2\pi a^2 \sin\theta' d\theta'}{[(z-a\cos\theta')^2 + (\varepsilon_{perp}/\varepsilon_{\parallel})a^2\sin^2\theta']^{1/2}}$$

$$= \frac{a\sigma}{2\varepsilon_{\parallel}} \int_{-a}^{a} \frac{dz'}{[(z'-z)^2 + (\varepsilon_{perp}/\varepsilon_{\parallel})(a^2-z'^2)]^{1/2}}, \quad (16)$$

since $z' = a\cos\theta'$, $dz' = a\sin\theta' d\theta'$ and $r_{\parallel}' = a\sin\theta'$. Completing the square in the denominator of the integrand and keeping only terms of first order in $\varepsilon_{perp}/\varepsilon_{\parallel}$, we get

$$\phi \approx \frac{a\sigma}{2\varepsilon_{\parallel}} \int_{-a}^{a} \frac{dz'}{[(z'-z)^2 + (\varepsilon_{perp}/\varepsilon_{\parallel})(a^2-z^2)]^{1/2}}$$

$$= \frac{a\sigma}{2\varepsilon_{\parallel}} \sum_{\pm} \text{arcsinh}\left[\frac{(a\mp z)^{1/2}}{(a\pm z)^{1/2}} \frac{\varepsilon_{\parallel}^{1/2}}{\varepsilon_{perp}^{1/2}}\right]. \quad (17)$$

Therefore, the self-energy is given by

$$U = (1/2)\int_0^{\pi} 2\pi a^2 \sin\theta d\theta \sigma\phi = \pi a\sigma \int_{-a}^{a} dz\phi = \frac{Q^2}{16\pi a^2 \varepsilon_{\parallel}} \int_{-a}^{a} dz\, \text{arcsinh}\left[\left(\frac{a-z}{a+z}\right)^{1/2}\left(\frac{\varepsilon_{\parallel}}{\varepsilon_{perp}}\right)^{1/2}\right]. \quad (18)$$

Then, substituting $\varepsilon_{\parallel} = 81\varepsilon_0$, $\varepsilon_{perp} = 2.1\varepsilon_0$, as well as several values obtained from simulations reported in Refs. 4 and 21, and performing the integral numerically in Eq. (18), we obtain The values for $U$ listed in Table I.

**Table I**

| $\varepsilon_\parallel$ | $\varepsilon_{perp}$ | $U[Q^2/(8\pi\varepsilon_\parallel a)]$ |
|---|---|---|
| $81\varepsilon_0$ | $2.1\varepsilon_0$ | 2.54 |
| $70\varepsilon_0$ | $2.1\varepsilon_0$ | 2.47 |
| $70\varepsilon_0$ | $\varepsilon_0$ | 2.83 |

Table I gives values found from Eq. (18) for U for $\varepsilon_\parallel = 81\varepsilon_0$, $\varepsilon_{perp} = 2.1\varepsilon_0$ and for several sample values from simulations reported in Refs. 4 and 21. .

For $\varepsilon_\parallel = 81\varepsilon_0$, $\varepsilon_{perp} = 2.1\varepsilon_0$ and for $a = 1.16 \times 10^{-10} m$ (the ionic radius of a sodium ion), $U = 3.12 \times 10^{-10} J$, and for $a = 1.67 \times 10^{-10} m$ (the ionic radius of a chloride ion), $U = 2.17 \times 10^{-10} J$. The solvation energy barrier (i.e., the difference between the self-energy within a distance *H* of the wall and its value beyond *H* from the wall) is given by

$$\Delta U = 1.54 \frac{Q^2}{8\pi\varepsilon_\parallel a}. \qquad (19)$$

For $a = 1.16 \times 10^{-10} m$ (the ionic radius of a sodium ion), $\Delta U = 1.90 \times 10^{-20} J = 4.74 k_B T_{room}$, and for $a = 1.67 \times 10^{-10} m$ (the ionic radius of a chloride ion), $\Delta U = 1.32 \times 10^{-20} J = 3.29 k_B T_{room}$.

Let us now examine the error introduced by setting $\vec{r}_\parallel = 0$ in the above integrals. The correct integral for $\vec{r}_\parallel \neq 0$ involves calculating the potential on a ring of radius $r_\parallel$ whose center is located at $z$ due to a ring of charge of radius $r_\parallel'$ whose center is located at $z'$. Therefore, the error introduced by setting $\vec{r}_\parallel = 0$ is proportional to the difference between the value of

$$I = \int_0^\infty dk J_0(kr_\parallel) J_0(kr_\parallel') \exp[k(\varepsilon_\parallel/\varepsilon_{perp})(z'-z)], \qquad (20)$$

which is proportional to the interaction of the two rings described above[22], for $r_\parallel = 0$ and for $r_\parallel = (a^2 - z^2)^{1/2}$. Values of *I* for several values of $z$ and $z'$ are shown in table II below:

**Table II**

| z | z' | I for $r_\parallel = (a^2 - z^2)^{1/2}$ | I for $r_\parallel = 0$ |
|---|---|---|---|
| 0.2a | -0.8a | 0.1584 | 0.1603 |
| 0.2a | -0.2a | 0.3554 | 0.3745 |
| 0.5a | -0.5a | 0.1580 | 0.1596 |
| 0.9a | -0.9a | 0.0894 | 0.0893 |

| 0.8a | -0.8a | 0.1005 | 0.1003 |

Table II: This table shows values of the integral *I* in Eq. (20) for the self-energy calculation, for several values of $z$ and $z'$, for both $\vec{r}_\parallel = 0$ and $r_\parallel = (a^2 - z^2)^{1/2}$.

Let us now calculate the image potential for an ion with a nonzero radius by direct integration. Consider a classical grounded metal wall located at $z = 0$. Consider the uniformly charged spherical shell model for an ion of radius *a* and charge Q, whose center is at $z = z_0$. Then, the boundary conditions on the metal's surface will be satisfied by adding to the electrical potential of the ion the potential due to an image which is a spherical ion of radius *a* and charge -Q whose center is located at $z = -z_0$. The image potential is the interaction energy between the ion and its image. To calculate it, we must calculate the electrical potential due to the ion at a point on the surface of the image with coordinates $(\vec{r}_\parallel, z)$, where $\vec{r}_\parallel = (x, y)$, which is given by

$$\phi = \frac{\sigma}{4\pi\varepsilon_\parallel} \int dA' \frac{1}{[(z'-z)^2 + (\varepsilon_{perp}/\varepsilon_\parallel)^{1/2} |\vec{r}_\parallel' - \vec{r}_\parallel|^2]^{1/2}}, \quad (21)$$

where dA' is an element of surface area on the ion. The calculation is illustrated in Fig. 3.

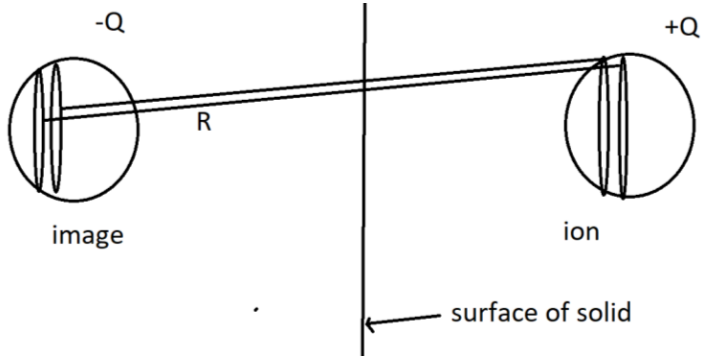

Fig. 3: The calculation in Eq. (21) is illustrated, where $R = [(z'-z)^2 + (\varepsilon_{perp}/\varepsilon_\parallel) |\vec{r}_\parallel' - \vec{r}_\parallel|^2]^{1/2}$.

Again, for simplicity we will set $\vec{r}_\parallel = 0$. The integral is now written as follows using spherical coordinates, $z' = z_0 + a\cos\theta'$, where $z' = z_0$ is the center of the ion:

$$\phi = \frac{\sigma}{4\pi\varepsilon_\parallel} \int 2\pi a^2 \sin^2\theta' d\theta' \frac{1}{[(z_0 + a\cos\theta' - z)^2 + (\varepsilon_{perp}/\varepsilon_\parallel)(a^2 - a^2\cos^2\theta')]^{1/2}}. \quad (22)$$

This can be rewritten as

$$\phi = \frac{\sigma a}{2\varepsilon_\parallel} \int_{-a}^{a} dz'' \frac{1}{[(z'' + z_0 - z)^2 - (\varepsilon_{perp}/\varepsilon_\parallel)(z''^2 - a^2)]^{1/2}}, \quad (23)$$

where $z'' = a\cos\theta'$. Completing the square and neglecting terms of higher than first order in $(\varepsilon_{perp}/\varepsilon_\parallel)$, the integral becomes

$$\phi = \frac{\sigma a}{2\varepsilon_\parallel}\int_{-a}^{a}\frac{dz''}{[(z''+z_0'')^2 - (\varepsilon_{perp}/\varepsilon_\parallel)(z_0''^2 - a^2)]^{1/2}} \tag{24}$$

where $z_0'' = z_0 - z$. Performing the integral, we obtain

$$\phi = \frac{\sigma a}{2\varepsilon_\parallel}\int_{-a}^{a}dz''\frac{1}{[(z''+z_0'')^2 - A^2]^{1/2}} = \frac{\sigma a}{2\varepsilon_\parallel}[\cosh\theta_+ - \cosh\theta_-], \tag{25}$$

where $A = (\varepsilon_{perp}/\varepsilon_\parallel)^{1/2}[(z-z_0)^2 - a^2]^{1/2}$ and

$$\theta_\pm = \left(\frac{z_0''\pm a}{[z_0''^2 - a^2]^{1/2}}\right)\left(\frac{\varepsilon_\parallel}{\varepsilon_{perp}}\right)^{1/2}.$$

Integrating z over the image we obtain for the image potential energy

$$U_{image} = -\frac{Q^2}{16\pi\varepsilon_\parallel a}\int_{2\bar{z}_0 - 1}^{2\bar{z}_0 + 1}d\bar{z}_0''[\cosh\theta_+ - \cosh\theta_-], \tag{26}$$

where $\bar{z}_0 = z_0/a$, $\bar{z}_0'' = z_0''/a$. Performing the integral in Eq. (26) numerically for $z_0 = 1.03a$, we obtain for the minimum value of the image interaction (i.e., its value for $z_0 = a$) for several values of $\varepsilon_\parallel$ and $\varepsilon_{perp}$, which are given in table III below. We chose $z_0 = 1.03a$ because the approximation of cutting off the expansion in $\varepsilon_{perp}/\varepsilon_\parallel$ breaks down at precisely $z_0 = a$.

**Table III**

| $\varepsilon_\parallel$ | $\varepsilon_{perp}$ | $U_{image}[Q^2/(8\pi\varepsilon_\parallel a)]$ |
|---|---|---|
| $81\varepsilon_0$ | $2.1\varepsilon_0$ | -1.30 |
| $70\varepsilon_0$ | $2.1\varepsilon_0$ | -1.31 |
| $70\varepsilon_0$ | $\varepsilon_0$ | -1.28 |

Table III gives values found from Eq. (26) for $U_{image}$ for $\varepsilon_\parallel = 81\varepsilon_0$, $\varepsilon_{perp} = 2.1\varepsilon_0$ and for several sample values from simulations reported in Refs. 4 and 21.      .

Assuming $\varepsilon_\| = 81\varepsilon_0$, $\varepsilon_{perp} = 2.1\varepsilon_0$ we obtain for $a = 1.16 \times 10^{-10} m$ (the ionic radius of a sodium ion), $U_{image} = -3.07 k_B T_{room}$, and for $a = 1.67 \times 10^{-10} m$ (the ionic radius of a chloride ion), $U_{image} = -2.13 k_B T_{room}$. Hence, the sum of the self-energy and the image potential at the wall is given by

$$U_{tot} = U + U_{image} = 1.25 \frac{Q^2}{8\pi\varepsilon_\| a}. \quad (27)$$

This implies that if salt water is made to flow between two metallic plates that are less than 2*H* apart, the salt concentration at the wall will be a factor of $\exp(-U_{image}/k_B T_{room}) = 12.5$ larger for the sodium ions and 8.41 larger for the chloride ions than it is when the ions are outside of the image potential well. This is the case because as an ion moves away from one of the walls, it encounters a second wall before its self-energy drops off to its bulk water value. If the walls are further apart, however, the ions will prefer to be a distance greater than *H* from the walls, where tables I and III and Eq. (27) show $U_{tot}$ to be smaller than its value at the walls. Thus, the ions would prefer to be concentrated midway between the walls. This will be illustrated in Fig. 5 below.

Let us again examine the error introduced by setting $\vec{r}_\| = 0$ in the above integral. The correct integral for $\vec{r}_\| \neq 0$ involves calculating the potential on a ring of radius $r_\|$ whose center is located at $z$ due to a ring of charge of radius $r_\|'$ whose center is located at $z'$. The error committed by setting $\vec{r}_\| = 0$ is proportional to the difference between *I*, defined in Eq. (20) for $r_\| = 0$, $r_\| = [a^2 - (z - 2z_0)^2]^{1/2}$. It is given for several values of $z$ and $z'$ in table IV:

**Table IV**

| z | z' | I for $r_\| = [a^2 - (z-2z_0)^2]^{1/2}$ | I for $r_\| = 0$ |
|---|---|---|---|
| 0.2a | -0.8a | 0.0536 | 0.0536 |
| 0.2a | -0.2a | 0.0668 | 0.0670 |
| 0.5a | -0.5a | 0.0536 | 0.0536 |
| 0.9a | -0.9a | 0.0424 | 0.0424 |
| 0.8a | -0.8a | 0.0447 | 0.0447 |

Table IV: This table shows values of the integral *I* in Eq. (20) for several values of *z* and *z'* for both $\vec{r}_\| = 0$ and $r_\| = [a^2 - (z - 2z_0)^2]^{1/2}$. Its role in the image potential calculation is described above Eq. (20).

Again, we see that the error committed by setting $\vec{r}_\| = 0$ is very small.

From Eq. (A10) in Appendix A, with $\kappa_{TF} = 0$, it is also clear that this same argument for the image potential can be used for the case of a wall made from a dielectric material of permittivity $\varepsilon_2 > (\varepsilon_\| \varepsilon_{perp})^{1/2}$ if we use for the image charge $-Q[\varepsilon_2 - (\varepsilon_\| \varepsilon_{perp})^{1/2}]/[\varepsilon_2 + (\varepsilon_\| \varepsilon_{perp})^{1/2}]$.

The discussion of ion solvation in this section is based on the Born approximation, which assumes a local dielectric constant. Theoretical studies of solvation based on a non-local dielectric constant and including more than one mode of polarization show that the Born approximation is an overestimate of the magnitude of the solvation energy[20,23-34].

We should also consider image charges resulting from the interface at a distance $H$ from the wall, where the dielectric constant of the water normal to the wall increases from 2.1 near the wall to 81 beyond a distance $H$ from the wall, in addition to the images from the surface at $z = 0$. There will be a series of multiple image charge terms reflecting the images of each of the image charges in the wall at $z = 0$ and the interface at $z = H$, located at $n(H \pm z_0)$, where $n$ is an integer[4]. These multiple images will be small for two reasons, and hence they will be neglected. First, it is more likely that the permittivity of water will change gradually as one moves from the wall to a distance greater than $H$ from the wall, and second, as we will discuss in section IV, there is screening due to the other ions in the solution, which will make the contribution of the higher order image charges fall off very rapidly with increasing $n$.

## IV. Self-Energy and Image Potential with Screening

There is experimental evidence that the static dielectric constant of an ionic solution decreases significantly as the ion concentration increases[23,34]. This suggests that the solvation of the ions decreases as the ion concentration increases. There have been many theoretical treatments that predict and confirm this result[23-34]. The presence of screening charge is expected to reduce the net charge on the sphere used to model the ion. One way to visualize this is to consider what would happen in the limit as the screening length is made to approach the ion's radius. The net charge on the surface of a spherical shell model for the ion (fixed charge plus polarization charge) would then be zero, and hence, there would be no net charge to interact with itself, resulting in the self-energy being zero. In this section we will discuss a simple model for screening that illustrates how screening reduces the self-energy.

Since Debye-Huckel theory is not accurate for small screening lengths, a theory of screening proposed by Nordholm, which agrees with Montecarlo calculations at high ion density at which Debye-Huckel theory breaks down[35], will be used. In this approach for a scalar dielectric constant, the screening charge density for a spherically symmetric charge distribution of total charge $Q$ centered around the origin is given by

$$\rho(r) = -Qn_B, \qquad a \leq r \leq h,$$
$$\rho(r) = -Qn_B \frac{he^{-K(r-h)}}{r}, \qquad r \geq h \qquad (28)$$

where $n_B$ is the number of ions of each charge per unit volume. The inverse Debye-Huckel screening length $K$ is given by

$$K = \left(\frac{4\pi n_B Q^2}{\varepsilon k_B T}\right)^{1/2} = 4\pi(\ell_B n_B)^{1/2} \qquad (29)$$

where $\ell_B = Q^2/(4\pi\varepsilon k_B T)$ is the Bjerrum length in MKS units, and $h$ is determined by the requirement that

$$\int_a^\infty 4\pi r^2 dr \rho(r) = -Q, \qquad (30)$$

which gives the following equation to solve for $h$: $x^3 + 3x^2 + 3x - A = 0$, where $x = Kh$ and $A = K^3[a^3 + 3/(4\pi n_B)]$. Making the substitution $x = v - 1$, the equation reduces to $v^3 = 1 + A$, and hence,

$$Kh = \left[1 + K^3(a^3 + 3/4\pi n_B)\right]^{1/3} - 1. \qquad (31)$$

For large $n_B$, $K^3 a^3 \gg 1$, which gives $h \approx [3/(4\pi n_B) + a^3]^{1/3}$. From Eq. (29) and using $\varepsilon = 81$, and hence, $\ell_B = 7.1\times 10^{-10}\,m$ we find that the large $n_B$ criterion is satisfied for $n_B \gg 7.48\times 10^{26}\,m^{-3}$, for $a = 1.10\times 10^{-10}\,m$ (sodium ions), and $3.25\times 10^{26}\,m^{-3}$, for $a = 1.67\times 10^{-10}\,m$ (chloride ions). For smaller values of the dielectric constant, $\ell_B$ will be even larger, and hence, the above inequality will likely be satisfied even for smaller vales of $n_B$. Thus, we are justified in using this simplified version of Nordholm's model for high salt concentrations. In this limit, the model reduces to a sphere of radius $h$, containing a screening charge density $n_B$ between this sphere and the surface of the spherical ion. Then, the amount of screening charge within this sphere (but outside of the ion) is given by

$$-Q n_B \left(\frac{4\pi(h^3 - a^3)}{3}\right) \approx -Q. \qquad (32)$$

Then within this simplified model, the electric displacement field is only nonzero within the sphere of radius $h$ and is given for a medium with a scalar permittivity by

$$D = \frac{Q}{4\pi r^2}\left(1 - \frac{r^3 - a^3}{h^3 - a^3}\right)\theta(r-a)\theta(h-r), \qquad (33)$$

where $\theta(x) = 0$ for $x < 1$ and 1 for $x > 1$, and similarly for the E-field. Therefore, the self-energy is given by

$$U = (1/2)\int_a^h 4\pi r^2 dr ED = (1/2)\int_a^h 4\pi r^2 dr E^2 = \frac{Q^2}{8\pi\varepsilon(h^3 - a^3)^2}\left(\frac{h^6}{a} + h^3 a^2 - \frac{9h^5 + a^5}{5}\right). \qquad (34)$$

This expression for $U$ includes both the work done to assemble the ion's charge and the screening charge from infinity because

$$U = (1/2)\int d^3r[\rho(\vec{r}) - \rho_s(\vec{r})]\phi = (1/2)\int d^3r(\nabla \cdot \vec{D})\phi = (1/2)\int d^3r \vec{D} \cdot \vec{E} \qquad (35)$$

where $\rho_s(\vec{r})$ is the screening charge and $\phi$ is the electrostatic potential.

The electric fields for a point charge located at $z = z_0$ near a wall (i.e., z<H), where the permittivity is a tensor, are given by Eqs. (10) and (11). In order to simplify the equations, we will again assume that the center of the ion is located at z=0 instead of $z = z_0$. As discussed in section III, this is the electric field due to a point charge located at the origin, but it is also equal to the electric field due to a charged spherical shell of total fixed charge $Q$, but with an angular dependence of the charge distribution on its surface, given by Eq. (12). We will use this model to illustrate the effect of screening because it can be studied analytically. The effect of screening on the self-energy should be similar for a uniformly charged spherical shell model for an ion. In order to effectively screen the field due to the charged shell, the density of ions of one charge, $n_B$ in Eq. (32) should be replaced by an ion density that has the angular distribution of the screening charge that matches the angular distribution of the fixed charge on the ion's surface, which is given by Eq. (12), because it is expected that the screening charge will arrange itself so as to completely screen the charge distribution on the ion's surface. Therefore $n_B$ should be replaced by

$$n_s = \left(\frac{\varepsilon_\parallel}{\varepsilon_{perp}}\right)^{1/2} \frac{n_B}{[(\varepsilon_\parallel / \varepsilon_{perp})\cos^2\theta + \sin^2\theta]^{3/2}}. \qquad (37)$$

. Then, Eq. (32) is replaced by

$$2\pi \int_a^h r^2 dr \int_0^\pi \sin\theta d\theta Q n_s = Q, \qquad (38)$$

with $n_s$ given by Eq. (37), which from Eqs. (37,38), reduces to Eq. (32). The screened electric fields in the Nordhom theory of screening are just Eqs. (10) and (11) multiplied by

$$\left[1 - \frac{r^3 - a^3}{h^3 - a^3}\right], \qquad (39)$$

and when these screened fields are substituted in the expression for the self-energy,

$$U = (1/2)\int d^3r \vec{E} \cdot \vec{D}, \qquad (40)$$

the self-energy becomes Eq. (14) multiplied by

$$\Gamma = [(h/a)^3 - 1]^{-2}[(h/a)^6 + (h/a)^3 - 1.8(h/a)^5 - 0.2] \qquad (41)$$

In the small salt concentration limit when $h/a \gg 1$, Eq. (41) becomes 1. In Fig. 4, $\Gamma$ is plotted as a function of $n_B a^3$ from $n_B a^3 = 10^{-5}$ to $n_B a^3 = 0.03$.

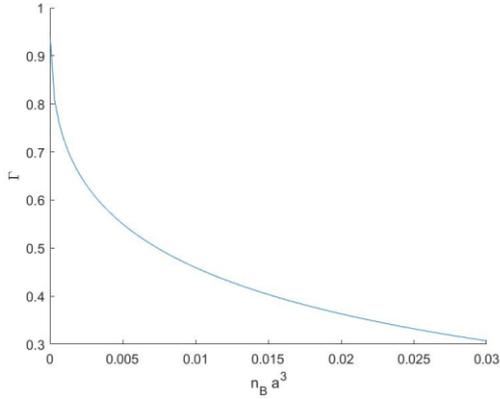

Fig. 4: The dimensionless quantity $\Gamma$ is plotted as a function of the dimensionless quantity $n_B a^3$.

The table below shows several values of $\Gamma$, for the sodium ion, which has a radius $a = 1.16 \times 10^{-10} m$, and the chloride ion, which has a radius $a = 1.67 \times 10^{-10} m$.

**Table V**

| $n_B$ | $\Gamma$ for a sodium ion | $\Gamma$ for a chloride ion |
|---|---|---|
| $10^{26} m^{-3}$ | 0.853 | 0.777 |
| $3.65 \times 10^{26} m^{-3}$ | 0.774 | 0.662 |
| $2 \times 10^{27} m^{-3}$ | 0.624 | 0.486 |
| $3 \times 10^{27} m^{-3}$ | 0.577 | 0.412 |
| $3.65 \times 10^{27} m^{-3}$ | 0.554 | 0.389 |

Table V: $\Gamma$ is given for several values of $n_B$, including $3.65 \times 10^{26} m^{-3}$, the salt concentration of sea water and $3.65 \times 10^{27} m^{-3}$, the salt concentration at the solubility limit of sodium chloride.

If we assume that a comparable reduction of the self-energy will occur for the uniformly charged spherical shell model for an ion described by Eqs. (15-18), we find that the self-energy at salt concentrations close to the solubility limit could be reduced to a value that is smaller than the image potential energy given in table III. This implies that there will be a negative total potential [i.e., $U_{tot}$ as defined in Eq. (27)] minimum at the wall experienced by the salt ions. For an ion at the wall, the image potential should not be significantly screened, and hence, should be approximately equal to the value given in table III. Since the self-energy at $z > H$ is positive, the self-energy plus image potential of an ion at the bottom of the potential well at the wall is lower than this energy for $z > H$, and hence, the ions will prefer to be at the wall. This is illustrated in Fig. 5, which shows the potential energy $v(z) = v_1(z) + v_2(z)$, where

$$v_1(z) = \frac{\Gamma U}{1 + \exp(-bz)} \qquad (42a)$$

represents the self-energy. The specific functional form was chosen because it describes qualitatively the behavior expected as a result of the position dependence of the dielectric constant observed near a wall in Ref. 19. The actual functional dependence of the self-energy on $z$ is not known. The self-energy near the wall $U$ is given in table I, and the screening factor $\Gamma$ is defined in Eq. (41). The potential

$$v_2(z) = \frac{U_{image}}{1 + c(z-d)^2} , \qquad (42b)$$

represents the potential well chosen to represent the image potential, where the minimum of this potential is located at $d = 0.5 \times 10^{-10} m$, and the other parameter values are given in the figure caption. This form is the one used in Ref. 15 to represent the image potential.

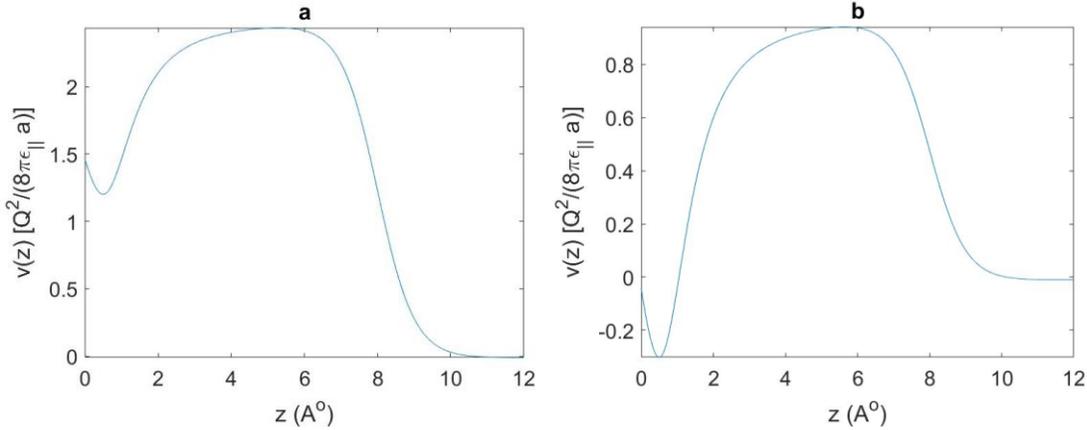

Fig. 5: Plots of $v(z)$ with $b = 2 \times 10^{10} m^{-1}$, $c = 10^{10} m^{-2}$, $U_{image} = -1.3 Q^2 / (8\pi \varepsilon_\parallel a)$ and (a) $U = 2.5 Q^2 / (8\pi \varepsilon_\parallel a)$, the value of the self-energy near the wall without screening, which comes from table I for $\varepsilon_\parallel = 81$, $\varepsilon_{perp} = 2.1$, to represent the case in which the self-energy is greater than the depth of the image potential well, and (b) $U = 0.998 Q^2 / (8\pi \varepsilon_\parallel a)$, the self-energy near the wall with screening factor $\Gamma = 0.399$ to represent a case in which self-energy near the wall is less than the depth of the image potential well.

The salt ions acted on by an self-energy and an image potential have a Boltzmann distribution. This follows from the minimization of a free energy consisting of the sum of the image potential, the self-energy and the entropy of mixing, as illustrated in appendix B and in Ref. 15 for ions acted on by an image potential. Since the ionic radius of the sodium ion is smaller than of that of the chloride ion, sodium ions will have a larger value of both the self-energy and the image potential. This will lead to a charge distribution near the wall.

## V. Thermal Activation over the Solvation Energy Barrier

The rate at which ions diffuse across the solvation energy barrier per unit area is crudely given by

$$J = (v_T / 2 \cdot 3^{1/2}) n e^{-\Delta U / k_B T_{room}} \qquad (43)$$

where $v_T$ is the rms (root mean square) thermal velocity of the ions, which is of the order of $10^2 m/s$ and $n$ is the ion density more than a distance $H$ from the wall. This is based on the idea that the mean velocity component towards the barrier is given by $(1/2) <v_z^2>^{1/2} = [1/(2 \cdot 3^{1/2})] <v^2>^{1/2}$. The thermal velocity $v_T = <v^2>^{1/2}$ was estimated by thinking of the ions in the liquid as being almost as close to neighboring water molecules or other ions as they would be in a solid. The Debye frequency in a solid is of the order of $10^{12} s^{-1}$ and the near neighbor distance is of the order of $10^{-10} m$. Since the mean velocity is of the order of the product of these two quantities, we obtain a mean velocity magnitude of $10^2 m/s$. The classical statistical mechanical expression for $v_T$, namely $(1/2) m v_T^2 = (3/2) k_B T$ gives 5.61X10² m/s for sodium ions and 4.52X10² m/s for the chloride atoms, but there was no way to anticipate a-priory that the classical treatment, which neglects quantum mechanics, would give the correct order of magnitude. Then, if we take $n = 7.3 \times 10^{26} ions/m^3$, the ion density for sea water, we find that for a barrier of $\Delta U = 3.29 k_B T_{room}$, the value for the unscreened solvation potential for chloride ions, $J = 0.783 \times 10^{27} ions/m^2 s$. Then the time that it takes for enough chloride ions to diffuse across the barrier and give rise to a concentration of $n_f = 7.3 \times 10^{27} ions/m^3$ at the wall, the ion density at the solubility limit, in a potential well of width of the order of $2 \times 10^{-10} m$ at the wall is given by $t = n_f a / J \approx 1.86 \times 10^{-9} s$. Here, $n_f a$ is equal to the number of ions that must flow through unit area in order to for the density of ions in the image potential well to be equal to $n_f$. The image potential well at a metallic wall actually has a width of the order of $\kappa_{TF}^{-1}$[18], the electronic screening length inside the metal, which for a good metal is of the order of $a$. Another way to think of this is that a fraction $\exp(-\Delta U / k_B T_{room})$ of the ions lies above the potential barrier. The number of these ions per unit area that are moving towards the image charge potential well per unit time is equal to $(1/2) n <v_z> \exp(-\Delta U / k_B T_{room}) = J$. The number of ions that have entered an image potential well of width of the order of $a$ in a time $t$ per unit area of the well is given by $Jt = n_f a$.

## VI. Conclusions

It has been shown that at low ion concentrations, at which ionic screening of the electrical potential acting on each ion is not important, the potential energy of an ion due its electrical image potential near a metallic wall is overcome by the solvation energy barrier (i.e., the difference between the self-energy near the wall and at a distance greater than $H=0.75nm$ from the wall). Since the sum of the self-energy and the image potential energy within a distance $H$ from the wall is higher than the self-energy at a distance greater than $H$, the ions will prefer to remain at a distance greater than $H$ from the wall. On the other hand, it was argued that at ion concentrations greater than that of sea water, the screening by the ions can reduce the solvation energy by a sufficient amount so that there is a potential minimum at the wall which is lower than the energy of an ion located more than $H$ from the wall, which should result in a high concentration of the salt at the wall. This suggests that the concentration of salt ions near a wall, as a result of the electrical image potential due to the wall, is more likely to be observed at salt concentration between that of sea water and the solubility limit of salt, where the salt ions occupy a large fraction of the volume of the solution. For example, for sodium chloride dissolved in water at its solubility limit, the diameter of a salt ion, even without including the hydration shell, is nearly 65% of the width of the volume available to each ion. This corresponds to the ions occupying 14% of the volume of the solution, so that not too many water molecules can be near each ion.

A recent paper by Misra and Blankschtein[5] reports all atom molecular dynamics simulations of the interaction of several negative ions in water with a finite sheet of graphene. They find that the presence of water significantly reduces the attractive interaction between each ion and the graphene from what it would be in a vacuum. The large reduction of the attraction between some of the ions and their model for undoped semimetallic graphene might be specific to that model, considering that attraction of ions to other surfaces is not found to be so severely reduced in other simulations[36]. Nevertheless, Misra and Blankschtein's results are qualitatively consistent with the results presented in this paper in the low ion density limit, although the mechanism for the interaction of the ions studied with the graphene might be different from the electrical image potential considered here.

**Acknowledgements**

**Appendix A**

**Solutions of the Poisson-Boltzmann Equation at a Metallic and a Dielectric Surface**

The potential for a point charge in a medium with a tensor permittivity is the solution to Poisson's equation in the region $z>0$, where there is a tensor permittivity described in section III in the paper,

$$\varepsilon_{perp}\frac{\partial^2 \phi}{\partial z^2} + \varepsilon_\parallel \left( \frac{\partial^2 \phi}{\partial x^2} + \frac{\partial^2 \phi}{\partial y^2} \right) = -Q\delta(\vec{r}-\vec{r}_0), \tag{A1}$$

with a point charge $Q$ located at $\vec{r}_0 = z_0 \hat{z}$, which is Eq. (7) in the main text. The region with $z<0$ will be assumed to have a scalar permittivity $\varepsilon_2$. Let us write $\phi$ in terms of its Fourier transform on the $x$ and $y$ coordinates,

$$\phi = (2\pi)^{-2} \int d^2 q e^{i(q_x x + q_y y)} \overline{\phi}(z,\vec{q}), \tag{A2}$$

where the solution to the differential equation obtained when Eq. (A2) is substituted in Eq. (A1) is

$$\bar{\phi} = \frac{Q}{2q(\varepsilon_{perp}\varepsilon_{\parallel})^{1/2}} e^{-q(\varepsilon_{\parallel}/\varepsilon_{perp})^{1/2}|z-z_0|} + Ae^{-q(\varepsilon_{\parallel}/\varepsilon_{perp})^{1/2} z}, \tag{A3}$$

for $z > 0$, where *A* is a constant, which satisfies the boundary condition that it vanishes at $z \to \infty$ [4,5,10]. If the region with z<0 is metallic, $\bar{\phi}$ satisfies

$$(\partial^2 z / \partial z^2 - \kappa_{TF}^2)\bar{\phi} = 0, \tag{A4}$$

where $\kappa_{TF}$ is the inverse Thomas-Fermi screening length of the metal, whose solution is

$$\bar{\phi} = Be^{\chi z}, \tag{A5}$$

where *B* is a constant and where $\chi = (q^2 + \kappa_{TF}^2)^{1/2}$, which satisfies the boundary condition that it vanishes as $z \to -\infty$. At the interface at z=0, the above solutions must be continuous (because of the requirement of continuity of the transverse E-field at the interface), giving

$$\frac{Q}{2q(\varepsilon_{perp}\varepsilon_{\parallel})^{1/2}} e^{-q(\varepsilon_{\parallel}/\varepsilon_{perp})^{1/2} z_0} + A = B. \tag{A6}$$

Continuity of the component of the displacement or D-field normal to the interface requires that

$$\varepsilon_{perp} \partial \phi(z > 0)/\partial z = \varepsilon_2 \partial \phi(z < 0)/\partial z \tag{A7}$$

as $z \to 0$, which gives

$$\varepsilon_{perp} \frac{Q}{4\pi(\varepsilon_{perp}\varepsilon_{\parallel})^{1/2}} \left(\frac{\varepsilon_{\parallel}}{\varepsilon_{perp}}\right)^{1/2} e^{-q\left(\frac{\varepsilon_{\parallel}}{\varepsilon_{perp}}\right)^{1/2} z_0} - \varepsilon_{perp} \left(\frac{\varepsilon_{\parallel}}{\varepsilon_{perp}}\right)^{1/2} A = \varepsilon_2 \chi B. \tag{A8}$$

The solution of Eqs. (A6) and (A8) is

$$A = \frac{Q}{2q(\varepsilon_{perp}\varepsilon_{\parallel})^{1/2}} \frac{(\varepsilon_{perp}\varepsilon_{\parallel})^{1/2} q - \varepsilon_2 \chi}{(\varepsilon_{perp}\varepsilon_{\parallel})^{1/2} q + \varepsilon_2 \chi} e^{-q(\varepsilon_{\parallel}/\varepsilon_{perp})^{1/2} z_0}$$

$$B = \frac{Qe^{-q(\varepsilon_{\parallel}/\varepsilon_{perp})^{1/2} z_0}}{\varepsilon_2(\chi/q) + (\varepsilon_{\parallel}/\varepsilon_{perp})^{1/2}}. \tag{A9}$$

The image potential energy is given by

$$QV_{imag} = Q\int \frac{d^2q}{(2\pi)^2} Ae^{-q(\varepsilon_{\parallel}/\varepsilon_{perp})^{1/2} z_0} = \frac{Q^2}{2(\varepsilon_{\parallel}\varepsilon_{perp})^{1/2}} \int \frac{d^2q}{(2\pi)^2} \frac{(\varepsilon_{perp}\varepsilon_{\parallel})^{1/2} q - \varepsilon_2 \chi}{(\varepsilon_{perp}\varepsilon_{\parallel})^{1/2} q + \varepsilon_2 \chi} \frac{e^{-2q(\varepsilon_{\parallel}/\varepsilon_{perp})^{1/2} z_0}}{q}. \tag{A10}$$

For small $z_0$ compared to $\kappa_{TF}^{-1}$, large q dominates the integral, and hence,

$$QV_{image} \approx -\frac{Q^2}{8\pi\varepsilon_\| z_0} \frac{\varepsilon_2 - (\varepsilon_\| \varepsilon_{perp})^{1/2}}{\varepsilon_2 + (\varepsilon_\| \varepsilon_{perp})^{1/2}}, \quad (A11a)$$

and for large $z_0$ small q dominates the integral, and hence,

$$QV_{image} \approx -\frac{Q^2}{4\pi\varepsilon_\| z_0}. \quad (A11b)$$

Since we have required that $z_0$ be restricted to values greater than *a*, Eq. (A11b) gives the minimum value of the image potential. The potential minimum, however, more likely occurs at $z_0 \approx \kappa_{TF}^{-1}$ [18], giving a potential minimum of the order of

$$QV_{image} \approx -\frac{Q^2 \kappa_{TF}}{4\pi\varepsilon_\|}. \quad (A12)$$

For a grounded classical metal for which the potential is zero on the interface, for z>0,

$$A = -\frac{Q}{2q(\varepsilon_{perp}\varepsilon_\|)^{1/2}} e^{-q(\varepsilon_\|/\varepsilon_{perp})^{1/2} z_0} \quad (A13)$$

which also gives

$$QV_{image} \approx -\frac{Q^2}{4\pi\varepsilon_\| z_0}. \quad (A14)$$

If the material located at z<0 is a dielectric, we set $\kappa_{TF} = 0$.

**Appendix B**

**A Simple Derivation of the Distribution of Salt Ions Acted on by an Electrical Image and a Solvation Potential**

The equilibrium distribution of the ions is determined by minimizing the free energy of the solution, which consists of the entropy of mixing and the potential energy of the ions. A common way to treat the entropy of mixing is to divide the volume of the solution into boxes, where each box has a volume $v_0$ of the order of the volume of a water molecule or a salt ion[37,38]. Therefore, the entropy of mixing is given by

$$S = k_B \sum_z \ln\left[\frac{(n_w(z)+n_s(z))!}{n_w(z)!n_s(z)!}\right] \approx k_B \left[\sum_z n_s(z) \ln\frac{n_w(z)+n_s(z)}{n_s(z)} + n_w(z) \ln\frac{n_w(z)+n_s(z)}{n_w(z)}\right], \quad (A1)$$

where $n_s(z), n_w(z)$ are respectively the number of salt ions and the number of water molecules located in the plane made up of boxes whose centers are located a distance $z$ from a wall. The Gibbs free energy is given by

$$G = G_0 + \sum_z n_s(z) U_{tot}(z) - TS, \qquad (A2)$$

where $G_0$ is the part of the Gibbs free energy that does not depend on the distribution of salt ions among the water molecules and $U_{tot}(z)$ is the total energy of an ion located in the plane of boxes a distance $z$ from the wall. In the model treated in this paper, it consists of the sum of the image potential and the self-energy of the ion, which for simplicity in this appendix we will choose to be the same for ions of both charges. The salt ion chemical potential is therefore given by

$$\mu_s(z) = \frac{\partial G}{\partial n_s(z)} = k_B T \sum_z \ln \frac{n_s(z)}{n_w(z)} + U_{tot}(z), \qquad (A3)$$

where we have made the assumption that $n_s(z) + n_w(z)$ is constant. (I.e., we are requiring that the solution is incompressible.) The equilibrium condition is

$$0 = \mu_s(z + \Delta z) - \mu_s(z) \approx -\left[ k_B T \frac{1}{x(z)} \frac{dx(z)}{dz} + \frac{dU_{tot}(z)}{dz} \right] \Delta z, \qquad (A4)$$

where $\Delta z$ is the distance between the successive values of $z$ and $x(z) = n_s(z)/n_w(z) \approx v_0 \rho_i(z)$. Substituting this in Eq. (A4) and solving the resulting differential equation, be obtain

$$\rho_i(z) = \rho_i(\infty) \exp\left[ -\frac{U_{tot}(z) - U_{tot}(\infty)}{k_B T} \right], \qquad (A5)$$

i.e., a Boltzmann distribution for the ions. Hence, the total potential acting on the ions alone determines the ions' spatial distribution.

## Data Availability

The data that supports the findings of this study are available within the article. They are either found in the tables or may be calculated form the equations contained in the article.

## Author Contribution Statement

J. B. Sokoloff is the sole author. All of the work reported in this manuscript was performed by him.

## References

1. P. M. Biesheuvel, S. Porada, M. Levi and M. Z. Bazant, J. Solid State Electrochemistry 18, 1365-1376 (2014).
2. J. Comtet, A. Nigues, V. Kaiser, B. Coasne, L. Bocquet and A. Siria, Nature Materials 16, 634-639 (2017).
3. R. Wang and Z.-G. Wang, Phys. Rev. Lett. 112, 136101 (2014).
4. Philip Loche, Cihan Ayaz, Alexander Schlaich, Douwe Jan Bonthuis and Roland R. Netz, J. Phys. Chem. Lett. 9, 6463 (2018)